\documentclass[journal]{IEEEtran}

\usepackage{amsmath,amssymb,amsfonts}
\usepackage[T1]{fontenc}
\usepackage[utf8]{inputenc}
\usepackage{graphicx}
\usepackage[export]{adjustbox}
\usepackage{booktabs}
\usepackage{multirow}
\usepackage{array}
\usepackage{float}
\usepackage{cite}
\usepackage{url}
\usepackage[unicode]{hyperref}
\graphicspath{{mathpix_assets/}}
\DeclareUnicodeCharacter{2713}{\checkmark}
\DeclareUnicodeCharacter{00D7}{\ensuremath{\times}}

\title{Dissipativity-Based Multiport Stability Root-Cause Identification and Mitigation for Solid-State Transformers}
\author{Xiangyu Meng, Dong Xie, Hongjian Lin, Chunxu Lin, Xinglai Ge, Zhigang Liu}

\begin{document}
\maketitle

Abstract-For solid-state transformers (SSTs) in highpower grid-connected applications, improperly designed control loops can excite strong inherent AC-DC port coupling, leading to low-frequency oscillation issues, especially under weak grid conditions. To address this problem, this article establishes a multiport admittance matrix for the SST, encompassing its AC $d q$ axes and primary DC port, to characterize its inherent dynamics. Subsequently, a multiport dissipativity analysis is conducted to evaluate the robust stability of SST. By leveraging the decomposition of passivity conditions into distinct self- and couplingdissipativity indices, the specific root causes of instability are diagnosed. This framework reveals that a severe coupling-dissipativity failure, induced by the internal dynamics of the synchronization loop, is the dominant instability mechanism rather than a localized self-dissipativity issue. Guided by this diagnosis, a stabilizing controller featuring dynamics-free orthogonal signal reconstruction is designed to reshape the admittance characteristics of the SST. This enhancement specifically targets the identified coupling-dissipativity deficiencies, thereby resolving

the root cause of the instability. Finally, the stability analysis and the effectiveness of the enhancement strategy are validated on a down-scaled SST prototype. Experimental results demonstrate that the criterion accurately predicts the coupling-induced oscillations and that the enhanced controller guarantees stable operation under challenging weak-grid conditions.

Index Terms-Admittance modeling, multiport systems, solid-state transformers (SSTs), stability criterion, weak grid stability.

\section*{I. Introduction}
THE solid-state transformer (SST) is a key technology for modern power systems, offering high power density and efficiency for renewable integration and microgrids [1]. However, the dynamic interaction between power converters in a SST-enabled hybrid AC-DC grid can lead to DC bus voltage instability, which manifests as low-frequency oscillation (LFO), particularly when the system damping ratio is low [2]. Without proper damping, such oscillations can lead to severe overvoltages and overcurrents, compromising the safety and reliability of the system operation.

To investigate the underlying oscillation mechanisms, several stability analysis techniques are available. Conventional approaches often rely on eigenvalue analysis [3]; however, the efficacy of this method is contingent upon obtaining a precise statespace model of the entire system. Such a model is impractical for the application at hand when the internal parameters of the converter and the grid impedance are unknown. To circumvent these challenges, impedance-based methods have emerged as the prevailing framework for stability assessment [4], [5]. This approach characterizes the converter and the grid by their respective admittances and impedances at the point of connection [6], where stability is subsequently determined by applying the generalized Nyquist criterion to their ratio. However, this method faces challenges in systems with multiple paralleled converters, where the aggregation of impedance data from all interacting converters becomes a complex and computationally burdensome undertaking [7].

As a powerful alternative that does not require a precise grid model, passivity theory offers a cornerstone for stability analysis and robust control design. Originating from classical

circuit theory, it defines a system as passive if its energydissipative property holds across all frequencies [8]. However, practical stability analysis in power electronics focuses on a critical frequency range where interactions are most prominent. Therefore, this article adopts the more general concept of dissipativity to describe this practical, frequency-limited characteristic. Its adoption in power electronics facilitated the analysis of single-input single-output (SISO) converter models [9]. This philosophy is reflected in standards like EN50388 and has been used to prevent destabilizing interactions in gridconnected converters [10]. To address the coupled dynamics of three-phase systems, the theory was extended to multiinput multioutput (MIMO) frameworks [11], where dissipativity is assessed through the properties of an admittance or impedance matrix. This MIMO approach has been instrumental in analyzing the stability of multiparalleled converters [12], with passivity indices being introduced to quantify the passivity margin of the system. However, the improper selection of a passivity index for common nonminimum phase systems can lead to erroneous stability conclusions, particularly at low frequencies [13]. It has been shown that tuning methods aimed at enhancing dissipativity can be contradictory and that minimizing the nonpassive frequency range can paradoxically result in unstable designs [13].

In response to these challenges, an extended frequencydomain passivity theory was proposed [14]. By introducing a weighting matrix, this framework effectively leverages the latent passive properties of grid inductance to provide a more precise criterion for low-frequency stability analysis. However, this work concentrates on the AC port dynamics of the converter. Recent advancements have extended passivity-based concepts to analyze the robust stability of general DC-DC and AC-DC interlinking converters using all-port admittance models [15], [16], these general frameworks have not yet been specifically tailored to the complex dynamics of SST. SST is inherently a multiport energy hub managing simultaneous power flows between the AC grid, internal DC links, and various local DC or AC loads (e.g., DC microgrids, AC subgrids, and energy storage systems). Unlike standard three-phase converters, single-phase SSTs rely on internal virtual signal generation mechanisms to construct the $d q$-frame, introducing unique cross-coupling dynamics that remain underexplored in existing multiport studies. A two-port model fails to capture the critical cross-coupling dynamics between these ports, which can be a primary source of instability [17], [18]. Consequently, a more comprehensive multiport evaluation is required.

To address these challenges, this article proposes a comprehensive stability analysis and mitigation framework for single-phase SSTs. The main contributions are summarized as follows.

\begin{enumerate}
  \item A precise three-port admittance model for the CHBDAB SST is established and verified. This model accurately characterizes the inherent dynamics of the SST by capturing the cross-coupling effects induced by virtual signal generation in single-phase systems.
  \item Based on the self- and coupling-dissipativity indices, a coupling-induced instability mechanism is revealed. The
\end{enumerate}

\begin{figure}[H]
\begin{center}
  \includegraphics[alt={},max width=\columnwidth]{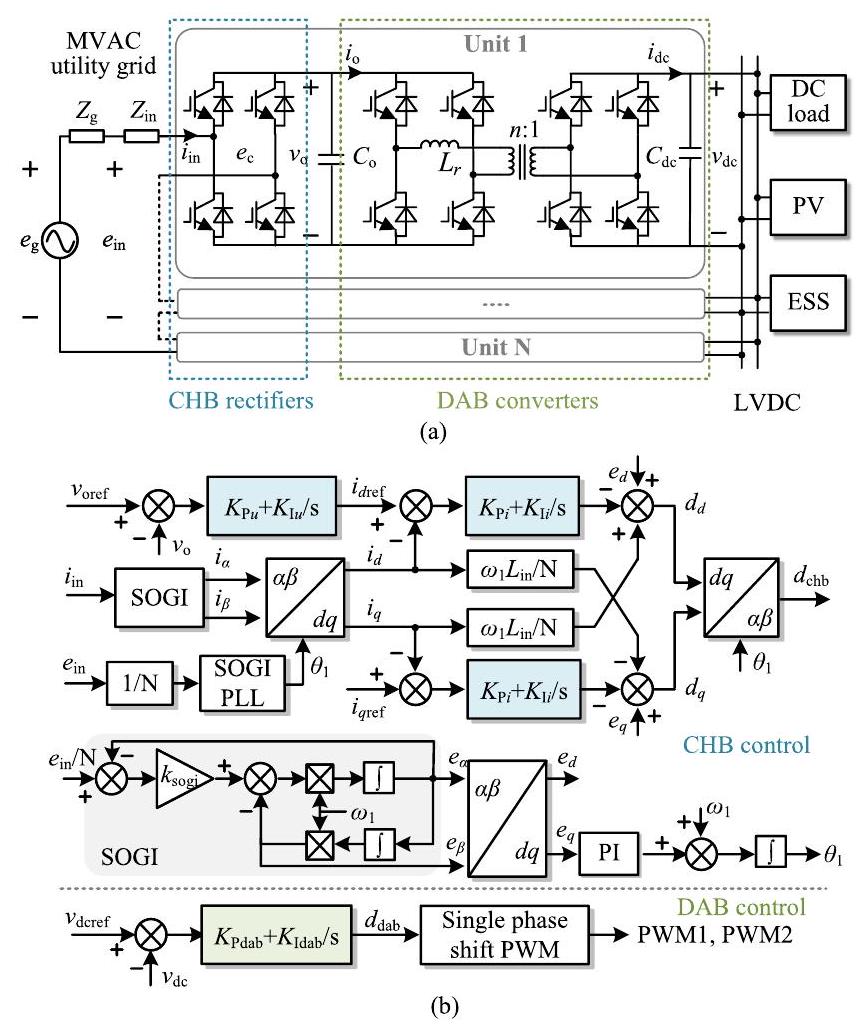}
\caption{System architecture of the CHB-DAB based SST. (a) Power stage topology. (b) Control system block diagram.}
\end{center}
\end{figure}

analysis quantitatively identifies that the internal dynamics of the SOGI-PLL act as a negative dissipativity source within the AC-DC coupling channel, providing a physical explanation for low-frequency oscillations that selfdissipativity indices fail to detect.\\
3) Guided by the diagnostic results, a stabilizing controller featuring dynamics-free orthogonal signal reconstruction (DF-OSR) is proposed. Unlike parameter retuning, which is often constrained by performance tradeoffs, this strategy structurally eliminates the delay-inducing filter dynamics from the coupling path, thereby restoring the system's passivity margin without compromising the dynamic bandwidth.\\
4) The proposed method is validated on a laboratory prototype. Experimental results confirm that the criterion not only accurately predicts the frequency of instability and that the enhanced controller guarantees stable operation under weak-grid conditions.

\section*{II. System Overview and SST Port Characterization Modeling}
\section*{A. Overview of SST Topology and Control}
The single-phase SST topology is depicted in Fig. 1. The architecture comprises two primary conversion stages: a CHB rectifier connected to the MV grid, and a subsequent isolated DC-DC stage. The CHB stage manages the high input voltage across $N$ cascaded cells, converting it to an intermediate DC voltage. This is followed by multiple DAB converter modules,\\
which provide high-frequency electrical isolation and regulate power flow to the local LVDC bus.

Regarding the modeling scope, two key simplifications are adopted. First, while a full AC-AC SST typically includes a downstream DC-AC inverter, this stage is modeled as an equivalent load on the LVDC bus to focus on grid-side interactions. Second, although applied in medium-voltage contexts, the CHB topology typically employs a phase-modular control architecture [19]. In this configuration, the three-phase system functions as three decoupled single-phase subsystems, thereby justifying the single-phase analysis presented in this work.

\section*{B. Small-Signal Transfer Functions of CHB and DAB Converters}
To derive the port characteristics of the entire SST, the smallsignal models of its constituent CHB and DAB stages are first established.

\begin{enumerate}
  \item DAB Converter: Based on the small-signal perturbation method around a steady-state operating point, the linearized input current $\Delta i_{\mathrm{o}}(s)$ and output current $\Delta i_{\mathrm{dc}}(s)$ of the DAB converter can be described as a function of perturbations in its control input $\Delta d_{\text{dab }}(s)$, input voltage $\Delta v_{\mathrm{o}}(s)$, and output voltage $\Delta v_{\mathrm{dc}}(s)$ [20]
\end{enumerate}

\[
\left\{\begin{array}{l}
\Delta i_{\mathrm{o}}(s)=G_{1 d}(s) \Delta d_{\mathrm{dab}}(s)+G_{1 u}(s) \Delta v_{\mathrm{dc}}(s)  \tag{1}\\
\Delta i_{\mathrm{dc}}(s)=G_{2 d}(s) \Delta d_{\mathrm{dab}}(s)+G_{2 u}(s) \Delta v_{\mathrm{o}}(s)
\end{array}\right.
\]

where $G_{1 d}(s), G_{1 u}(s), G_{2 d}(s)$, and $G_{2 u}(s)$ are the transfer functions derived from the state-space averaging model of the DAB converter.

As illustrated in the control diagram, the DAB converter employs a closed-loop voltage control to maintain a constant output voltage. The relationship between the output voltage error and the phase-shift ratio is governed by a PI controller, $G_{d v}(s)$

\begin{equation*}
\Delta d_{\mathrm{dab}}(s)=-G_{d v}(s) \Delta v_{\mathrm{dc}}(s) \tag{2}
\end{equation*}

where $G_{d v}(s)=K_{\text{Pdab }}+K_{\text{Idab }} / s$.\\
The nodal equation for the intermediate capacitor, considering the DAB input current as the disturbance, is given by

\begin{equation*}
s C_{\mathrm{o}} \Delta v_{\mathrm{o}}=-\Delta i_{\mathrm{o}} . \tag{3}
\end{equation*}

Additionally, by combining the (1), (2) and (3), the voltage transfer function $T_{\text{vo\_vdc }}(s)$ can be obtained as

\begin{equation*}
T_{v o \_v d c}(s)=\frac{\Delta v_{\mathrm{o}}}{\Delta v_{\mathrm{dc}}}=-\frac{G_{1 u}(s)-G_{1 d}(s) G_{d v}(s)}{s C_{\mathrm{o}}} . \tag{4}
\end{equation*}

\begin{enumerate}
  \setcounter{enumi}{1}
  \item CHB Converter: From Fig. 1(a), the AC -side dynamics in the $d q$-frame are expressed as
\end{enumerate}

\begin{equation*}
\Delta \boldsymbol{e}_{d q}(s)=\boldsymbol{Z}_{d q}(s) \Delta \boldsymbol{i}_{d q}(s)+N \Delta \boldsymbol{e}_{c d q}(s) \tag{5}
\end{equation*}

where $\Delta \boldsymbol{e}_{\mathrm{c} d q}$ is the CHB-side modulation voltage of cell, $\boldsymbol{Z}_{d q}(s)$ is the impedance matrix of the input filter, and $\Delta \boldsymbol{i}_{d q}$ and $\Delta \boldsymbol{e}_{d q}$ are the $d q$ components of AC-side current $i_{\text{in }}$ and voltage $e_{\text{in }}$ vectors, respectively.

The use of the $d q$ frame is necessitated by the inner loop control of CHB, which employs proportional-integral (PI) regulators on $d q$ current components to achieve zero steady-state\\
error. To align the plant model with this control structure, the physical single-phase impedance is transformed into the $d q$ domain. This is achieved by constructing a virtual orthogonal $\beta$-axis component to form a stationary $\alpha \beta$ system, followed by the Park transformation. This process introduces frequencydependent cross-coupling terms due to the frame rotation. The resulting impedance matrix $Z_{d q}(s)$ is given by

\[
Z_{d q}(s)=\left[\begin{array}{cc}
R_{\mathrm{in}}+s L_{\mathrm{in}} & -\omega_{1} L_{\mathrm{in}}  \tag{6}\\
\omega_{1} L_{\mathrm{in}} & R_{\mathrm{in}}+s L_{\mathrm{in}}
\end{array}\right]
\]

where $\omega_{1}$ is the fundamental grid frequency. This matrix characterizes the inherent coupling between the $d$ and $q$ axes induced by the frame rotation.

The coupling between the AC and DC sides is established through two fundamental relationships: the modulation equation and the power balance principle. In the small-signal domain, the converter-side modulation voltage and the intermediate DC-side current are expressed as

\begin{align*}
\Delta \boldsymbol{e}_{\mathrm{c} d q}(s) & =V_{o} \Delta \boldsymbol{d}_{d q}(s)+\boldsymbol{D}_{d q} \Delta v_{\mathrm{o}}(s)  \tag{7}\\
\Delta i_{\mathrm{o}}(s) & =\frac{1}{N V_{\mathrm{o}}}\left(\boldsymbol{E}_{d q}^{\mathrm{T}} \Delta \boldsymbol{i}_{d q}(s)+\boldsymbol{I}_{d q}^{\mathrm{T}} \Delta \boldsymbol{e}_{d q}(s)\right)-\frac{I_{\mathrm{o}}}{V_{\mathrm{o}}} \Delta v_{\mathrm{o}}(s) \tag{8}
\end{align*}

where $\Delta \boldsymbol{d}_{d q}$ is the duty cycle vector for cell, and $\Delta v_{\mathrm{o}}$ and $\Delta i_{\mathrm{o}}$ are the intermediate DC voltage and current, respectively.

The duty cycle vector, $\Delta \boldsymbol{d}_{d q}$, is generated by a comprehensive closed-loop control system. Considering the cascaded effects of the PLL, SOGI, current controller, and system delays, the duty cycle can be expressed as a function of the AC-side currents and voltages, and the current reference

\begin{align*}
\Delta \boldsymbol{d}_{d q}(s)= & \boldsymbol{G}_{d i 1}(s) \Delta \boldsymbol{i}_{d q, \text{ ref }}^{\mathrm{c}}(s)+\boldsymbol{G}_{d i 2}(s) \Delta \boldsymbol{i}_{d q}(s) \\
& +\boldsymbol{G}_{d e}(s) \Delta \boldsymbol{e}_{d q}(s) \tag{9}
\end{align*}

where the superscript "c" denotes variables defined within the $d q$ reference frame of the controller, which can deviate from the global reference frame of the system due to the dynamics of the PLL. The transfer function matrices $\boldsymbol{G}_{d i 1}, \boldsymbol{G}_{d i 2}$, and $\boldsymbol{G}_{d e}$ encapsulate the dynamics of the aforementioned control blocks, [5], [21].

Furthermore, the current reference itself is generated by the outer DC voltage control loop

\begin{equation*}
\Delta \boldsymbol{i}_{d q, \text{ ref }}^{c}(s)=-\boldsymbol{G}_{i v}(s) \Delta v_{\mathrm{o}}(s) \tag{10}
\end{equation*}

where $\boldsymbol{G}_{i v}(s)=\left[K_{\mathrm{P} u}+K_{\mathrm{I} u} / s \quad 0\right]^{T}$.\\
Then, based on the above process, the internal control variables and links the AC-side dynamics directly to the intermediate DC voltage can be derived as

\begin{align*}
\Delta \boldsymbol{e}_{\mathrm{c} d q}= & V_{\mathrm{o}} \boldsymbol{G}_{d e} \Delta \boldsymbol{e}_{d q}+V_{\mathrm{o}} \boldsymbol{G}_{d i 2} \Delta \boldsymbol{i}_{d q} \\
& +\left(V_{\mathrm{o}} \boldsymbol{G}_{d i 1} \boldsymbol{G}_{i v}+\boldsymbol{D}_{d q}\right) \Delta v_{\mathrm{o}} . \tag{11}
\end{align*}

In practical CHB implementations, circuit parameter tolerances (e.g., capacitance mismatch) require an active DC-link voltage balancing loop, which introduces asymmetric modulation indices among cells. However, this article employs a symmetric model for the stability analysis. As demonstrated in [22], the differential duty cycles summed across all modules approximate to zero, having negligible impact on the aggregate\\
port characteristics. Consequently, the symmetric model captures the dominant grid-side dynamics.

\section*{C. Multiport Admittance Model of the SST}
By integrating (4) and (11) into (5), the ac-side admittance model can be obtained as

\begin{align*}
& \underbrace{\left(-\boldsymbol{Z}_{d q}(s)-V_{\mathrm{o}} \boldsymbol{G}_{d i 2}\right)}_{A} \Delta \boldsymbol{i}_{d q} \\
& =\underbrace{\left(V_{\mathrm{o}} \boldsymbol{G}_{d e}-\boldsymbol{I}\right)}_{B} \Delta \boldsymbol{e}_{d q}+\underbrace{\binom{V_{\mathrm{o}} \boldsymbol{G}_{d i 1} \boldsymbol{G}_{i v} T_{v o \_v d c}}{+\boldsymbol{D}_{d q} T_{v o \_v d c}}}_{C} \Delta v_{\mathrm{dc}} \\
& \Rightarrow \Delta \boldsymbol{i}_{d q}=\underbrace{\boldsymbol{A}^{-1} \boldsymbol{B}}_{\boldsymbol{Y}_{d q}} \Delta \boldsymbol{e}_{d q}+\underbrace{\boldsymbol{A}^{-1} \boldsymbol{C}}_{\boldsymbol{Y}_{d q, \mathrm{dc}}} \Delta v_{\mathrm{dc}} \tag{12}
\end{align*}

This equation provides the first two rows of the multiport admittance matrix, namely the AC-side self-admittance matrix $\boldsymbol{Y}_{d q}$ and the DC-to-AC coupling admittance vector $\boldsymbol{Y}_{d q, \mathrm{dc}}$.

Assuming lossless operation, the total AC input power must equal the total DC output power. By linearizing the power balance around the steady-state operating point, an intrinsic relationship between all port variables is established as

\begin{equation*}
\boldsymbol{E}_{d q}^{\mathrm{T}} \Delta \boldsymbol{i}_{d q}(s)+\boldsymbol{I}_{d q}^{\mathrm{T}} \Delta \boldsymbol{e}_{d q}(s)=N\left(V_{\mathrm{dc}} \Delta i_{\mathrm{dc}}(s)+I_{\mathrm{dc}} \Delta v_{\mathrm{dc}}(s)\right) . \tag{13}
\end{equation*}

Furthermore, by substituting (12) into (13), one can solve for the remaining admittance terms

\begin{align*}
Y_{\mathrm{dc}, d}(s) & =\frac{1}{N V_{\mathrm{dc}}}\left(E_{d} Y_{d d}(s)+E_{q} Y_{q d}(s)+I_{d}\right)  \tag{14}\\
Y_{\mathrm{dc}, q}(s) & =\frac{1}{N V_{\mathrm{dc}}}\left(E_{d} Y_{d q}(s)+E_{q} Y_{q q}(s)+I_{q}\right)  \tag{15}\\
Y_{\mathrm{dc}}(s) & =\frac{1}{N V_{\mathrm{dc}}}\left(E_{d} Y_{d, \mathrm{dc}}(s)+E_{q} Y_{q, \mathrm{dc}}(s)-N I_{\mathrm{dc}}\right) \tag{16}
\end{align*}

These expressions demonstrate the AC-DC coupling characteristics and DC-port dynamics, thus completing the multiport model.

As illustrated by the Norton equivalent circuit in Fig. 2, the DC port is treated as one port, while the AC side splits naturally into the $d$ axis and the $q$ axis. The low-voltage DClink capacitance is treated as an external termination in this model, rather than an internal component of the admittance matrix. This boundary definition decouples the converter dynamics from the DC bus configuration. The small-signal behavior of the SST is characterized by applying an independent voltage perturbation $\mathbf{v}(s)=\left[\begin{array}{lll}\Delta e_{d}(s) & \Delta e_{q}(s) & \Delta v_{\mathrm{dc}}(s)\end{array}\right]$ at each port and measuring the resulting current responses $\mathbf{i}(s)= \left[\Delta i_{d}(s) \quad \Delta i_{q}(s) \quad \Delta i_{\mathrm{dc}}(s)\right]$. This leads to the definition of a $3 \times 3$ admittance matrix $\mathbf{Y}(s)$, which in the $d q$-frame can be written as

\[
\mathbf{Y}(s)=\left[\begin{array}{ccc}
Y_{d d}(s) & Y_{d q}(s) & Y_{d, \mathrm{dc}}(s)  \tag{17}\\
Y_{q d}(s) & Y_{q q}(s) & Y_{q, \mathrm{dc}}(s) \\
Y_{\mathrm{dc}, d}(s) & Y_{\mathrm{dc}, q}(s) & Y_{\mathrm{dc}}(s)
\end{array}\right] .
\]

Although the SST is inherently a linear time-periodic system, the proposed stability analysis is based on the $d q$ synchronous frame model. This transformation maps the fundamental frequency coupling effects into the constant cross-coupling

\begin{figure}[H]
\begin{center}
  \includegraphics[alt={},max width=\columnwidth]{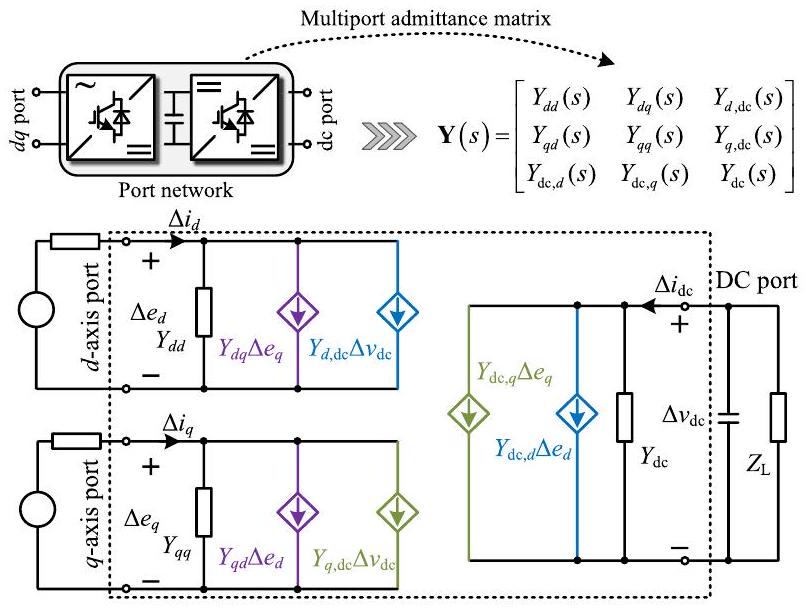}
\caption{Small-signal multiport admittance equivalent circuit of the SST. The model defines the self-admittance at each port and visually distinguishes the cross-coupling admittances: $d-q$ coupling (purple), $d$-dc coupling (blue), and $q$-dc coupling (green).}
\end{center}
\end{figure}

elements of the MIMO admittance matrix, thereby deriving an equivalent LTI model valid for small-signal analysis. This validates the application of the LTI passivity criterion within the control bandwidth of interest [21].

\section*{III. Multiport Dissipativity Criterion for Stability Assessment}
This section presents the multiport dissipativity criterion by the established admittance models in the previous section. Specifically, it enables evaluation of the self-dissipativity of the AC and DC ports individually, as well as the interactive dissipativity between them.

\section*{A. Multiport Dissipativity Condition for the SST}
The dissipativity of the multiport SST system, described by the admittance matrix $\mathbf{Y}(s)$, can be formally evaluated in the frequency domain. A linear time-invariant (LTI) system is passive if and only if its admittance matrix $\mathbf{Y}(s)$ is positive real. For a stable system, this condition is equivalent to requiring that the Hermitian part of the admittance matrix, $\mathbf{Q}(j \omega)$, is positive semidefinite for all frequencies $\omega$ [23]. The matrix $\mathbf{Q}(j \omega)$ is defined as

\begin{equation*}
\mathbf{Q}(j \omega)=\frac{1}{2}\left(\mathbf{Y}(j \omega)+\mathbf{Y}^{H}(j \omega)\right) \geq 0, \quad \forall \omega \in \mathbb{R} \tag{18}
\end{equation*}

where $\mathbf{Y}^{H}(j \omega)$ is the conjugate transpose (Hermitian) of $\mathbf{Y}(j \omega)$. However, this conventional criterion is known to be conservative when applied to grid-connected converters, particularly at low frequencies. To address this challenge, the extended frequency-domain passivity theory is proposed [14]. This theory generalizes the conventional criterion into a less conservative extended passivity condition by introducing a frequency-dependent weighting matrix $\mathbf{R}(\omega)$ [14]

\begin{align*}
\mathbf{Q}_{\mathrm{R}}(j \omega)= & \frac{1}{2}\left(\mathbf{R}^{H}(\omega) \mathbf{Y}(j \omega)\right. \\
& \left.+\mathbf{Y}^{H}(j \omega) \mathbf{R}(\omega)\right) \geq 0, \quad \forall \omega \in \mathbb{R} \tag{19}
\end{align*}

\section*{B. Decomposition of Dissipativity Properties}
To provide a more intuitive and powerful tool for analysis and controller design, the positive semidefinite condition in (19) is decomposed by evaluating the principal minors of the matrix $\mathbf{Q}_{\mathbf{R}}(j \omega)$. According to Sylvester's criterion for Hermitian matrices [24], $\mathbf{Q}_{\mathrm{R}}(j \omega)$ is positive semidefinite if and only if all of its principal minors are nonnegative. This approach allows us to define distinct "dissipativity properties" that isolate the contributions of self-admittances and cross-coupling admittances to the overall system stability. To make the calculation more explicit, we first define a weighted admittance matrix

\begin{align*}
\mathbf{Y}_{\mathrm{R}}(j \omega) & =\mathbf{R}^{H}(\omega) \mathbf{Y}(j \omega) \\
& =\left[\begin{array}{ccc}
Y_{\mathrm{R} d d}(s) & Y_{\mathrm{R} d q}(s) & Y_{\mathrm{R} d, \mathrm{dc}}(s) \\
Y_{\mathrm{R} q d}(s) & Y_{\mathrm{R} q q}(s) & Y_{\mathrm{R} q, \mathrm{dc}}(s) \\
Y_{\mathrm{Rdc}, d}(s) & Y_{\mathrm{Rdc}, q}(s) & Y_{\mathrm{Rdc}}(s)
\end{array}\right] \tag{20}
\end{align*}

where the weighting matrix $\mathbf{R}(\omega)=\operatorname{diag}\left(\mathbf{R}_{\mathrm{ac}}(\omega), 1\right)$ is the key component of the extended frequency-domain passivity theory [14]. It is a frequency-dependent matrix used to reduce the conservativeness of the standard passivity test. At low frequencies (e.g., $f<f_{1}$ ), which is the range of interest for LFOs, $\mathbf{R}_{\mathrm{ac}}(\omega)$ is set to a $2 \times 2$ rotation matrix $[0,1 ;-1,0]$. This rotation is a mathematical tool that effectively leverages the latent passive properties of grid inductance. A grid-connected converter often appears nonpassive (active) at its terminals at low frequencies. This rotation accounts for the inherent passivating effect of the series inductance of the grid, without needing to know the grid's specific impedance value. This makes the criterion much less conservative and more applicable to realworld systems. At high frequencies (e.g., $f \geq f_{1}$ ), $\mathbf{R}_{\mathrm{ac}}(\omega)$ is set to the $2 \times 2$ identity matrix $I$. In this range, the system dynamics are different, and the extended criterion (19) simply reduces to the conventional passivity criterion (18).

Physically, the order of the principal minors corresponds to the dimension of the energetic subsystem being evaluated: first-order minors ( $P_{d}, P_{q}, P_{\mathrm{dc}}$ ) represent the self-dissipativity of isolated ports, while higher-order minors characterize the coupling-dissipativity of interacting subsystems ( $P_{\mathrm{dc}, d}, P_{\mathrm{dc}, q}$ ) and the global stability ( $P_{\text{all }}$ ).

\begin{enumerate}
  \item Self-Dissipativity Properties: These properties ensure that each analytical port and the coupled AC-side subsystem are passive when considered in isolation.\\
a) DC-port self-dissipativity ( $P_{\mathrm{dc}}$ ): This is the real part of the DC-side admittance, ensuring the DC port itself is dissipative.
\end{enumerate}

\begin{equation*}
P_{\mathrm{dc}}(j \omega)=\operatorname{Re}\left\{Y_{\operatorname{Rdc}}(s)\right\} \geq 0 . \tag{21}
\end{equation*}

b) AC-side self-dissipativity ( $P_{d}, P_{q}, P_{d q}$ ): These three properties ensure the dissipativity of the AC-side $d q$ subsystem.

\begin{align*}
P_{d}(j \omega) & =\operatorname{Re}\left\{Y_{\mathrm{R} d d}(s)\right\} \geq 0  \tag{22}\\
P_{q}(j \omega) & =\operatorname{Re}\left\{Y_{\mathrm{R} q q}(s)\right\} \geq 0  \tag{23}\\
P_{d q}(j \omega) & =\operatorname{det}\left[\begin{array}{ll}
\operatorname{Re}\left\{Y_{\mathrm{R} d d}\right\} & \frac{Y_{\mathrm{R} d q}+\overline{Y_{\mathrm{R} q d}}}{2} \\
\frac{\overline{Y_{\mathrm{R} d q}+Y_{\mathrm{R} q d}}}{2} & \operatorname{Re}\left\{Y_{\mathrm{R} q q}\right\}
\end{array}\right] \geq 0 \tag{24}
\end{align*}

where $P_{d}$ and $P_{q}$ represent the dissipativity of individual axes, and $P_{d q}$ accounts for the interaction between the $d$ and $q$ axes.\\
2) Coupling-Dissipativity Properties: These properties address the dynamic coupling between the AC and DC ports, evaluating whether the energy exchange between them remains passive. Ensuring these conditions are met is crucial for preventing instabilities that arise purely from the interaction of otherwise stable subsystems.

\begin{enumerate}
  \item AC-DC port coupling ( $P_{\mathrm{dc}, d}, P_{\mathrm{dc}, q}$ ): These properties ensure that the interaction between the DC port and each individual AC axis remains passive.
\end{enumerate}

\begin{align*}
& P_{\mathrm{dc}, d}(j \omega)=\operatorname{det}\left[\begin{array}{cc}
\frac{\operatorname{Re}\left\{Y_{\mathrm{R} d d}\right\}}{\frac{Y_{\mathrm{R} d, \mathrm{dc}}+Y_{\mathrm{Rdc}, d}}{2}} & \frac{Y_{\mathrm{R} d, \mathrm{dc}}+\overline{Y_{\mathrm{Rdc}, d}}}{2} \\
\operatorname{Re}\left\{Y_{\mathrm{Rdc}}\right\}
\end{array}\right] \geq 0  \tag{25}\\
& P_{\mathrm{dc}, q}(j \omega)=\operatorname{det}\left[\begin{array}{cc}
\frac{\operatorname{Re}\left\{Y_{\mathrm{R} q q}\right\}}{\frac{Y_{\mathrm{R} q, \mathrm{dc}}+\overline{Y_{\mathrm{Rdc}, q}}}{2}} \\
\frac{Y_{\mathrm{R} q, \mathrm{dc}}+Y_{\mathrm{Rdc}, q}}{2} & \operatorname{Re}\left\{Y_{\mathrm{Rdc}}\right\}
\end{array}\right] \geq 0 \tag{26}
\end{align*}

\begin{enumerate}
  \setcounter{enumi}{2}
  \item All-Port Dissipativity Property: This final property ensures that the entire multiport system is passive when all interactions are considered simultaneously
\end{enumerate}

\begin{equation*}
P_{\mathrm{all}}(j \omega)=\operatorname{det}\left(\mathbf{Q}_{\mathrm{R}}(j \omega)\right) \geq 0 \tag{27}
\end{equation*}

By satisfying all seven of these conditions, particularly the coupling-dissipativity properties, robust stability of the SST can be ensured against any passive termination from the utility grid and connected loads. This framework provides a clear and systematic target for controller design, shifting the focus from simply shaping a single output impedance to actively managing the multiport dissipativity characteristics of the entire converter.

\section*{C. Dissipativity-Oriented Controller Design Workflow}
The proposed multiport dissipativity criterion enables a systematic, diagnostic-driven workflow for robust controller design. As illustrated in Fig. 3, this iterative procedure guides the development from initial specifications to a final controller validated for both multiport stability and dynamic performance. The workflow consists of four stages.

\begin{enumerate}
  \item Stage 1. Initialization and Baseline Design: The process begins by defining the system parameters and stability requirements of the SST. Crucially, this stage also involves setting the baseline dynamic performance metrics (e.g., desired bandwidth, transient response targets). A baseline controller, designed with conventional PI tuning to first satisfy these performance requirements, then serves as the benchmark for the subsequent dissipativity assessment. To avoid the conservatism of a fullspectrum analysis, the assessment is focused on a key frequency range, $\Omega$, which covers critical control bandwidths.
  \item Stage 2. Dissipativity Analysis: Next, the multiport admittance matrix, $\mathbf{Y}(s)$, is formulated per (17). The seven decomposed dissipativity properties are then calculated across $\Omega$ using (21)-(27). If all seven properties are nonnegative, the design is considered stable and proceeds to Stage 4. However, given the inherent conservatism of the criterion, minor violations (i.e.,
\end{enumerate}

\begin{figure}[H]
\begin{center}
  \includegraphics[alt={},max width=\columnwidth]{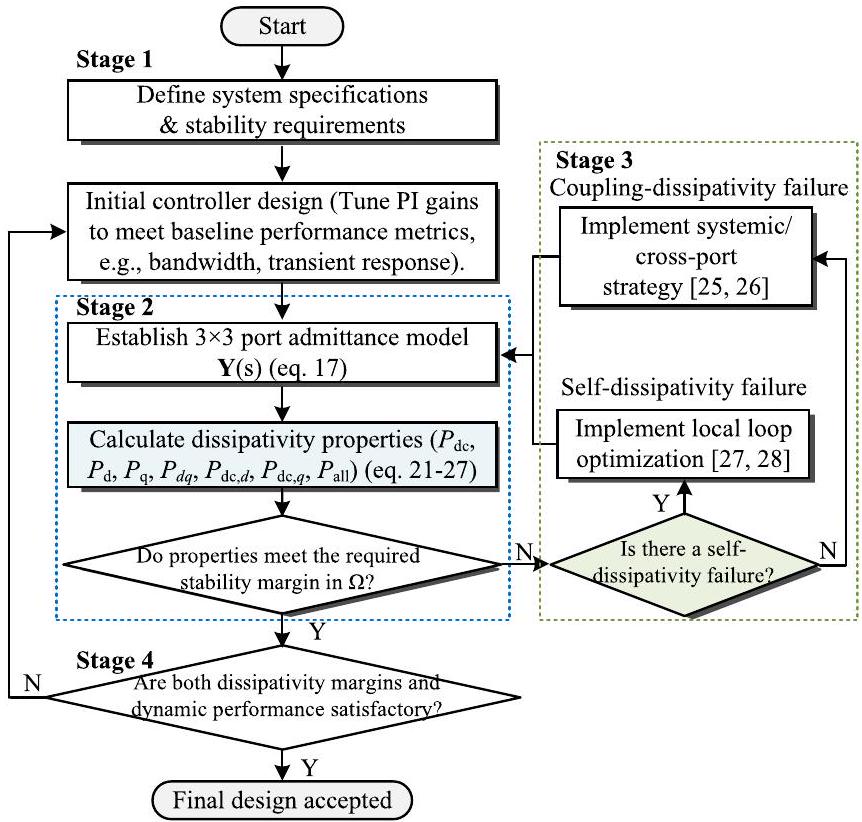}
\caption{Systematic design workflow based on the multiport dissipativity criterion.}
\end{center}
\end{figure}

small negative values over a narrow frequency band) may be deemed acceptable based on a predefined stability margin and engineering judgment. A significant violation, characterized by a large negative magnitude in a critical frequency sub-band, triggers the redesign process in Stage 3.\\
3) Stage 3. Diagnostic-Based Controller Redesign: This stage embodies the key advantage of the proposed methodology. Instead of trial-and-error, the redesign is guided by diagnosing the specific nature of the dissipativity failure. A hierarchical approach is taken.

\begin{enumerate}
  \item A self-dissipativity failure ( $P_{\mathrm{dc}}, P_{d}, P_{q}$, or $P_{d q}<0$ ) indicates a localized issue within a specific port or subsystem, often caused by poorly-tuned regulators or control delays. This is addressed with targeted methods such as PI gain tuning, active damping via derivative controllers [25], or multisampled PWM to reduce digital delays [26].
  \item A coupling-dissipativity failure ( $P_{\mathrm{dc}, d}, P_{\mathrm{dc}, q}$, or $P_{\mathrm{all}}<$ 0 ), occurring despite all self-dissipativity properties being met, points to a systemic issue from adverse port interactions. This requires more advanced strategies like feedforward decoupling or virtual impedance emulation [27], [28].\\
Following the implementation of a control enhancement, the workflow returns to Stage 2 to re-evaluate the dissipativity of the modified system.
  \item Stage 4. Final Validation and Tradeoff Analysis: A design satisfying the dissipativity criterion is not yet final. This stage validates the overall design against both stability and dynamic performance criteria. The key objective is to manage the inherent tradeoff between performance and stability: modifications made in Stage 3 to ensure dissipativity may degrade the dynamic performance (e.g., transient response) established in Stage 1.
\end{enumerate}

\section*{IV. Dissipativity-Based Analysis and Controller Enhancement}
This section demonstrates the application of the multiport dissipativity criterion established in Section III. The analysis begins by assessing a baseline SST controller to identify its dissipativity deficiencies. Then, based on diagnosis workflow, a targeted control enhancement strategy is developed.

\section*{A. Baseline Controller Design and Parameter Derivation}
The baseline controller parameters were derived using classical control methodologies [29]. The design objective was to meet standard frequency-domain specifications, such as bandwidth and phase margin.

\begin{enumerate}
  \item CHB Control Loop Tuning: The inner current loop of the CHB rectifier employs a PI regulator. Its parameter design follows the modulus optimum (MO) criterion to achieve a fast dynamic response while compensating for the input filter pole [29]. The proportional gain $K_{\mathrm{P} i}$ is determined by the desired current loop bandwidth $\alpha_{c}$ (set to 200 Hz to be well within $1 / 10$ of the switching frequency) and the input filter inductance $L_{\text{in }}$
\end{enumerate}

\begin{equation*}
K_{\mathrm{P} i}=\alpha_{c} L_{\mathrm{in}} . \tag{28}
\end{equation*}

The integral gain $K_{\mathrm{I} i}$ is selected to cancel the dominant pole of the plant ( $R_{\text{in }} / L_{\text{in }}$ ), yielding $K_{\mathrm{I} i}=K_{\mathrm{P} i} R_{\text{in }} / L_{\text{in }}$.

For the outer DC-link voltage loop, which regulates a plant dominated by the capacitor integrating characteristic, the symmetrical optimum (SO) criterion is adopted [29]. With a target voltage loop bandwidth of $f_{\mathrm{vc}}=10 \mathrm{~Hz}$, the proportional gain $K_{\mathrm{P} u}$ and integral gain $K_{\mathrm{I} u}$ are derived based on the DC-side capacitance $C_{\mathrm{o}}$

\begin{equation*}
K_{\mathrm{P} u} \approx C_{\mathrm{o}} \omega_{\mathrm{vc}}, \quad K_{\mathrm{I} u} \approx C_{\mathrm{o}} \omega_{\mathrm{vc}}^{2} . \tag{29}
\end{equation*}

Substituting the physical parameters from Table I ( $L_{\text{in }}=3.1 \mathrm{mH}, C_{\mathrm{o}}=2.8 \mathrm{mF}$ ), the calculated gains ensure sufficient phase margin for the decoupled loops.\\[0pt]
2) DAB Voltage Loop Tuning: Similarly, the DAB converter's output voltage regulation is tuned using the SO criterion [2], treating the DAB as a controlled current source charging the output capacitor $C_{\mathrm{dc}}$. The small-signal gain $G_{i d}$ is first derived from the phase-shift modulation characteristics:

\begin{equation*}
G_{i d} \approx \frac{v_{\mathrm{dc}}}{2 \pi f_{\mathrm{sw}, \mathrm{dab}} L_{\mathrm{r}}} . \tag{30}
\end{equation*}

The PI gains $K_{\text{Pdab }}$ and $K_{\text{Idab }}$ are then calculated to achieve a bandwidth of 35 Hz , ensuring fast load transient response.\\[0pt]
3) SOGI-PLL Parameter Tuning: Crucially, the SOGI-PLL parameters are designed to ensure robust grid synchronization. Based on the linearized small-signal model and SO principles [30], the SOGI gain $k_{\text{sogi }}$ is coupled to the target bandwidth $\omega_{\text{ff }}$. The design dictates

\begin{equation*}
k_{\mathrm{sogi}}=\frac{2}{\tau_{p} \omega_{\mathrm{ff}}} \tag{31}
\end{equation*}

where $\tau_{p}$ is a time constant derived from the design bandwidth [30]. For a target PLL bandwidth of 5 Hz , the theoretical calculation yields an optimal SOGI gain of $k_{\text{sogi }} \approx 0.5$.

\begin{table}[H]
\begin{center}
\caption{Main Hardware and Control Parameters of the Experimental SST Prototype}
\begin{tabular}{|l|l|l|}
\hline
Category & Parameter & Value \\
\hline
\multirow{6}{*}{System ratings} & Rated power ( $P_{\text{rated }}$ ) & 800 W \\
\hline
 & Grid voltage (rms) & 50 V \\
\hline
 & Grid frequency & 50 Hz \\
\hline
 & Nominal DC voltage ( $v_{\text{dcref }}$ ) & 80 V \\
\hline
 & Switching frequency (CHB) & 3 kHz \\
\hline
 & Switching frequency (DAB) & 10 kHz \\
\hline
\multirow{7}{*}{Hardware} & Grid inductance $\left(L_{\mathrm{g}}\right)$ & See Table II \\
\hline
 & Input filter inductance ( $L_{\text{in }}$ ) & 3.1 mH \\
\hline
 & Input filter resistance ( $R_{\text{in }}$ ) & $0.2 \Omega$ \\
\hline
 & CHB DC-link capacitor ( $C_{\mathrm{o}}$ ) & 2.8 mF \\
\hline
 & DAB leakage inductance ( $L_{\mathrm{r}}$ ) & 0.6 mH \\
\hline
 & Output DC capacitor ( $C_{\mathrm{dc}}$ ) & $420 \mu \mathrm{~F}$ \\
\hline
 & Number of modules ( $N$ ) & 3 \\
\hline
\multirow{6}{*}{CHB control} &  &  \\
\hline
 & Current loop bandwidth Current loop gains ( $K_{\mathrm{P} i}, K_{\mathrm{I} i}$ ) & 200 Hz (3.89, 250.9) \\
\hline
 &  &  \\
\hline
 & Voltage loop bandwidth Voltage loop gains ( $K_{\mathrm{P} u}, K_{\mathrm{I} u}$ ) & 10 Hz (0.17, 11) \\
\hline
 &  &  \\
\hline
 & PLL bandwidth SOGI gain & 5 Hz See Table II \\
\hline
\multirow{2}{*}{DAB control} & Voltage loop bandwidth & 35 Hz \\
\hline
 & Voltage loop gains ( $K_{\text{Pdab }}, K_{\text{Idab }}$ ) & (0.043, 9.58) \\
\hline
\end{tabular}
\end{center}
\end{table}

The full set of derived control parameters is listed in Table I. These theoretically parameters serve as the input for the subsequent multiport dissipativity analysis.\\[0pt]
4) Model Accuracy Verification: The frequency scanning method for MIMO systems was employed to measure the port admittance characteristics of the SST's detailed switching model [31]. To fully identify the $3 \times 3$ matrix, three independent perturbation sequences (injecting voltage at $\mathrm{AC}-d, \mathrm{AC}$ $q$, and DC ports, respectively) were applied. Fig. 4 compares the analytically derived admittance elements (solid lines) with the measured data (dots). The precise alignment across the frequency range confirms that the mathematical model accurately describes the system's coupling dynamics.

\section*{B. Dissipativity Assessment of the Initial Design}
The initial controller shown in Fig. 1(b), hereafter referred to as the "baseline controller", is designed in previous section. Its PI gains for the $d-q$ axis current loops and the DC voltage loop are tuned to achieve a fast transient response and minimal steady-state error. While this article uses conventional PI controllers to establish the baseline admittance model, the multiport dissipativity diagnostic framework is a general tool. It is equally applicable to analyzing SSTs that employ other control strategies, such as sliding mode control (SMC) or model predictive control (MPC). In such cases, the corresponding admittance matrix $\mathbf{Y}(s)$ would be derived based on the specific control law, and the same seven dissipativity indices would be calculated to diagnose the system's stability properties.

To quantify the stability risks under these conditions, the multiport dissipativity criterion is applied to the admittance model of the SST with the baseline controller. The frequency

\begin{figure}[H]
\begin{center}
  \includegraphics[alt={},max width=\columnwidth]{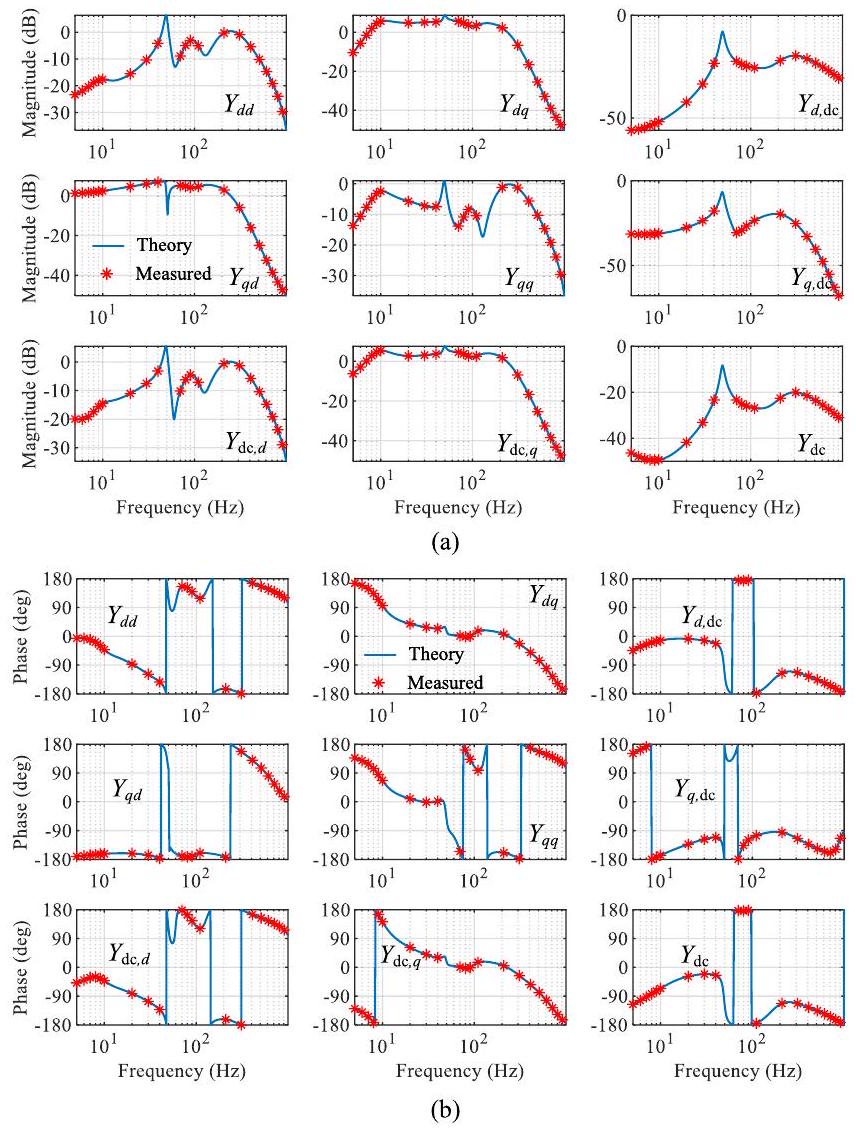}
\caption{Verification of the SST small-signal model: Comparison between the analytical admittance model (solid lines) and the frequency scanning measurements from the detailed switching model (markers).}
\end{center}
\end{figure}

responses of the seven decomposed dissipativity properties are calculated and plotted in Fig. 5. This analysis investigates two key aspects: the impact of the SOGI gain, denoted as $k_{i \text{ sogi }}$, on system stability, and the intrinsic dissipativity characteristics introduced by the active DAB stage.

The analysis first reveals the influence of the SOGI gain on the system low-frequency dissipativity. As depicted in Fig. 5(f) and (g), a clear trend is observable in the $10-50 \mathrm{~Hz}$ range. As $k_{i \text{ sogi }}$ is reduced, the indices for DC- $q$ coupling-dissipativity ( $P_{\mathrm{dc}, q}$ ) and all-port dissipativity ( $P_{\text{all }}$ ) decrease, indicating a erosion of the stability margin. This phenomenon can be attributed to the role of the SOGI within the grid synchronization loop. A lower $k_{\text{isogi }}$ value corresponds to reduced damping in the AC signal filtering, which can lead to oscillations in the estimated grid phase angle. These phase angle fluctuations propagate through the control system, adversely affecting the dynamic coupling between the $q$-axis and the DC-link voltage regulation, thereby degrading the $P_{\mathrm{dc}, q}$ property and, consequently, the overall system dissipativity $P_{\text{all }}$.

Furthermore, the analysis reveals the inherent destabilizing effect of the active DAB stage. A direct comparison between the active DAB case (solid lines) and the deactivated case (red dashed line) is revealed in the magnified view of Fig. 5(a). When the DAB is active, the DC-port self-dissipativity ( $P_{\mathrm{dc}}$ ) is negatively shifted, exhibiting a nondissipative characteristic

\begin{figure}[H]
\begin{center}
  \includegraphics[alt={},max width=\columnwidth]{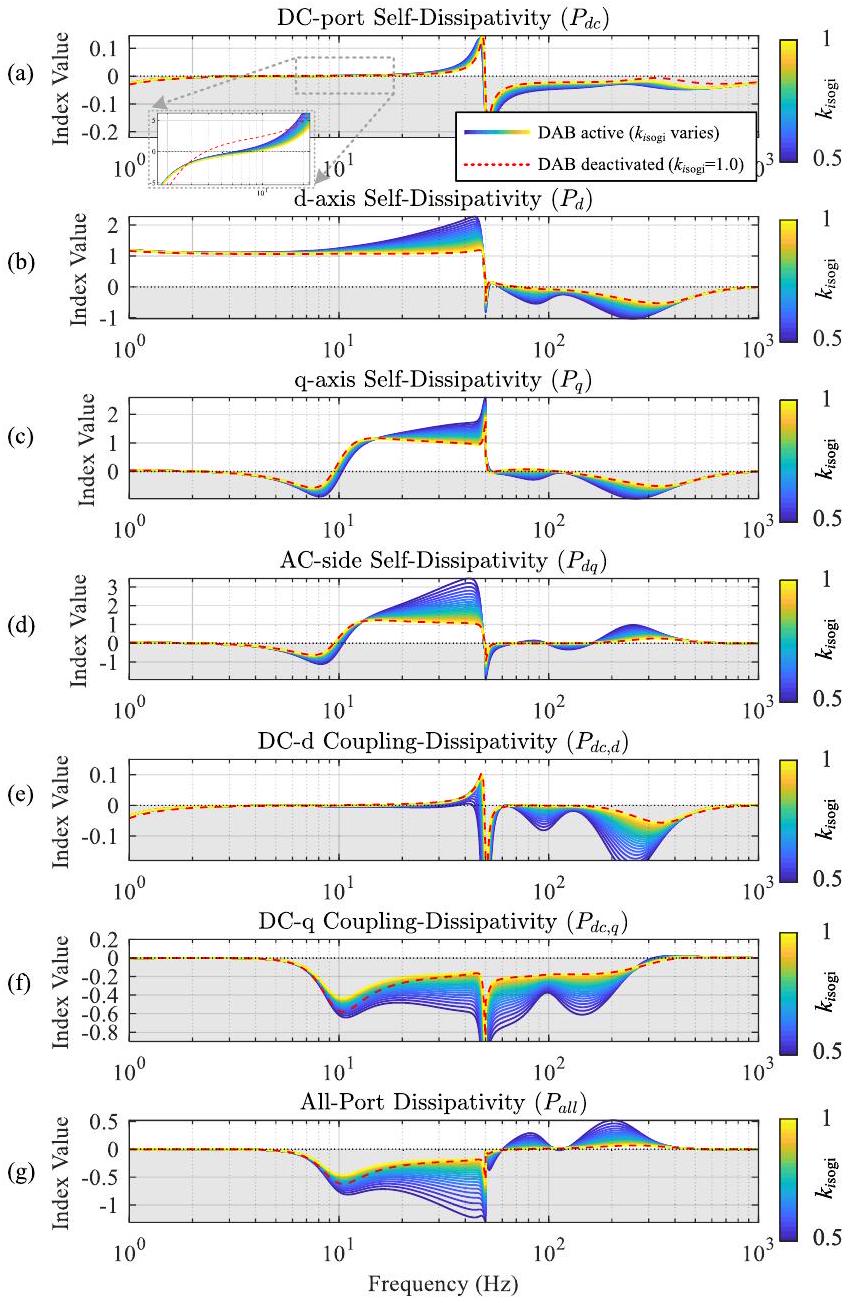}
\caption{Dissipativity properties of the SST with the baseline controller. (a)-(d) Self-dissipativity properties. (e) and (f) Coupling-dissipativity properties. (g) All-port dissipativity.}
\end{center}
\end{figure}

across the low-frequency spectrum. This is a direct consequence of the DAB's voltage control loop, which forces the stage to behave as a constant power load. CPLs are well-known for introducing a negative incremental impedance, which manifests as a negative dissipativity index, representing an intrinsic source of instability.

In summary, the dissipativity assessment of the baseline controller identifies two distinct stability vulnerabilities: 1) an intrinsic nondissipativity at the DC port caused by the CPL behavior of the DAB stage; and 2) a conditional instability risk linked to the AC-DC coupling channels, which is exacerbated by the improper tuning of the SOGI gain. This diagnosis underscores that achieving robust stability requires addressing not only the intrinsic properties of the load but also the dynamic interactions governed by the AC-side controllers.

\section*{C. Diagnostic-Based Controller Enhancement}
The analysis identifies the primary stability risk as a severe coupling-dissipativity failure-a systemic issue that simple PI

\begin{figure}[H]
\begin{center}
  \includegraphics[alt={},max width=\columnwidth]{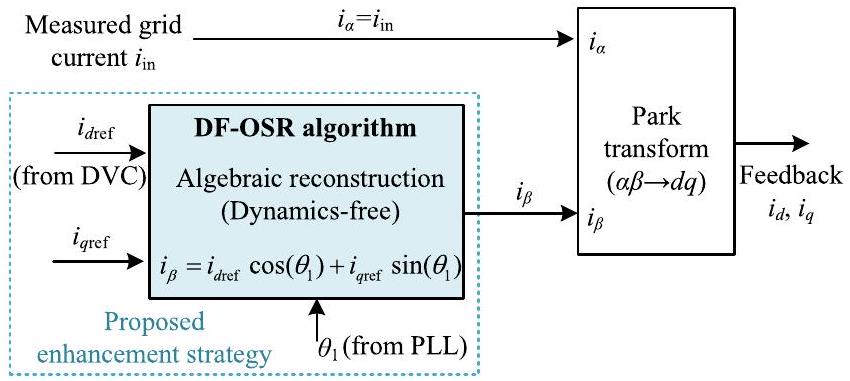}
\caption{Control implementation diagram of the proposed DF-OSR strategy.}
\end{center}
\end{figure}

tuning cannot resolve. Consequently, an advanced control strategy is necessary to reshape the admittance of SST and ensure multiport dissipativity.

To directly address the root cause of this adverse coupling, a controller enhancement strategy based on dynamics-free orthogonal signal reconstruction is implemented. This strategy alters how the quadrature current signal, which is essential for $d q$-frame control, is generated. Unlike conventional methods that use dynamic filters (e.g., SOGI) to estimate the quadrature signal from the measured current, this approach algebraically synthesizes the signal directly from the controller's command references.

Ad shown in Fig. 6, the principle is to apply the inverse Park transformation to the $d q$-frame current reference values ( $i_{d \text{ ref }}, i_{q \text{ ref }}$ ) to obtain the corresponding signals in the stationary $\beta$-axis current frame. The relationship is defined as

\begin{equation*}
i_{\beta}=i_{d \mathrm{ref}} \cos \left(\theta_{1}\right)+i_{q \mathrm{ref}} \sin \left(\theta_{1}\right) . \tag{32}
\end{equation*}

The reconstructed quadrature signal, $i_{\beta}(t)$, is then used in the control loop, replacing the output of a conventional dynamic filter. While the quadrature ( $\beta$-axis) signal is synthesized using this formula, the direct ( $\alpha$-axis) signal used for the Park transformation is the actual measured ac-side current ( $i_{\text{in }}$ ). This ensures that the control loop maintains feedback from the real system.

It is important to clarify the structural distinction regarding the SOGI usage in this enhanced scheme. The SOGI-PLL is retained solely in the voltage synchronization loop to provide the accurate phase angle $\theta_{1}$ required for the coordinate transformations in (32). In contrast, the SOGI dynamics within the current feedback loop—previously diagnosed as the root cause of the coupling-dissipativity failure ( $P_{\mathrm{dc}, q}<0$ )-are completely replaced by the proposed DF-OSR. This separation ensures that the system maintains robust grid synchronization without introducing the delay-induced negative damping associated with SOGI-based signal generation in the critical coupling path.

The primary benefit of this algebraic reconstruction is the complete elimination of the resonant dynamics and phase lag intrinsically associated with SOGI-based filters. This modification directly remedies the coupling-dissipativity failure diagnosed previously. The AC-DC coupling admittances (e.g., $\left.Y_{d, \mathrm{dc}}(s)\right)$ represent how a disturbance propagates from the DC port through the control system to the AC currents. A dynamic filter introduces its own undesirable phase characteristics into

\begin{figure}[H]
\begin{center}
  \includegraphics[alt={},max width=\columnwidth]{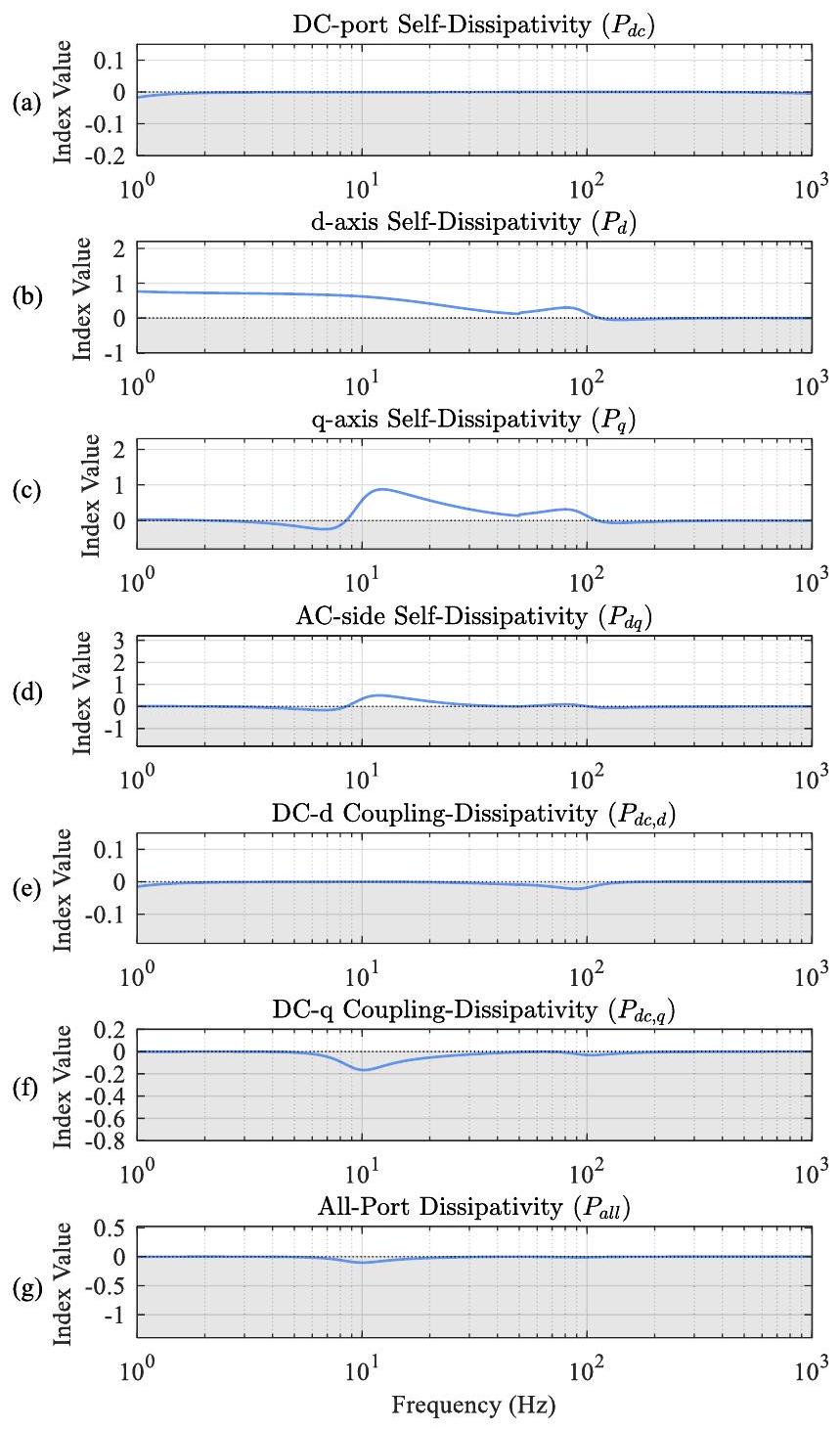}
\caption{Dissipativity properties of the SST with the DF-OSR controller. (a)-(d) Self-dissipativity properties. (e) and (f) Coupling-dissipativity properties. (g) All-port dissipativity.}
\end{center}
\end{figure}

this propagation path, degrading the stability margin of the coupling channel and leading to a nonpassive response. By providing a dynamics-free, instantaneous quadrature signal, the proposed enhancement creates a more ideal control platform, fundamentally improving the dynamic response of the coupling paths and mitigating the adverse interactions between the AC and DC ports.

\section*{D. Dissipativity Re-Assessment of the Enhanced Design}
To theoretically validate the effectiveness of the DF-OSR enhancement, the multiport dissipativity analysis is repeated for the SST system with the modified control structure.

The results presented in Fig. 7 demonstrate the enhanced dissipativity properties of the SST with the DF-OSR controller. The most significant improvement is observed in the coupling-dissipativity channels, previously identified as the primary source of instability. As shown in Fig. 7(e) and (f), the

\begin{figure}[H]
\begin{center}
  \includegraphics[alt={},max width=\columnwidth]{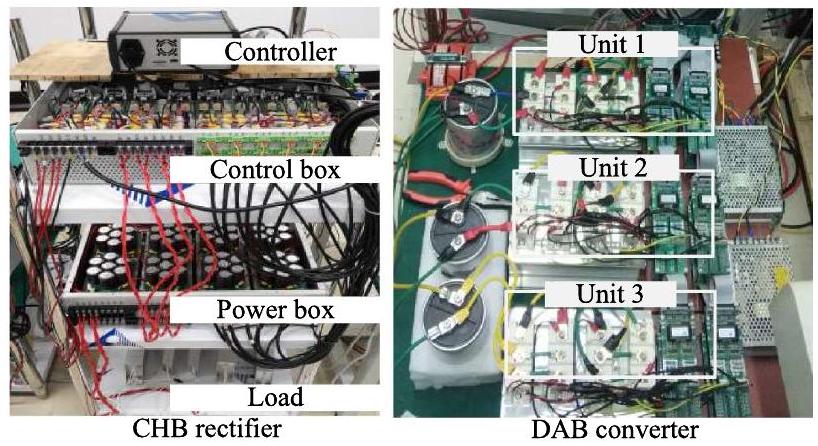}
\caption{Laboratory prototype of the SST-interfaced grid-connected system.}
\end{center}
\end{figure}

severe negative dips in both the DC- $d\left(P_{\mathrm{dc}, d}\right)$ and DC- $q\left(P_{\mathrm{dc}, q}\right)$ coupling indices are reduced. This confirms that the DF-OSR strategy effectively reduces the adverse dynamic interactions between the AC and DC ports.

Furthermore, the enhancement yields significant improvements in the self-dissipativity properties. While the intrinsic nondissipative characteristic of the active DAB remains, as evidenced by the slightly negative $P_{\mathrm{dc}}$ in Fig. 7(a), its detrimental impact on the overall system has been neutralized by the improvements elsewhere. More importantly, the AC-side self-dissipativity indices ( $P_{d}, P_{q}$, and $P_{d q}$ ) are now positive, showcasing the well-damped and stable behavior of the ACside controller.

\section*{V. Experimental Validation}
To verify the theoretical analysis and demonstrate the effectiveness of the proposed dissipativity-based design methodology, a down-scaled laboratory prototype of the SST-interfaced grid-connected system was developed. This section presents the experimental results.

\section*{A. Experimental Setup}
The experimental validation was performed on the laboratory prototype shown in Fig. 8, whose main hardware and control parameters are detailed in Table I. The grid emulation setup comprises a programmable AC source (Chroma). To emulate different grid stiffness conditions, a physical inductor ( $L_{\mathrm{g}}$ ) is connected in series between the AC source and the SST input. The capacitor $C_{\mathrm{g}}$ is connected in parallel with the series combination of the grid inductance and the AC source. The grid strength is quantified by the short circuit ratio (SCR), defined as follow [32]

$$
\mathrm{SCR}=\left(e_{\mathrm{in}}^{2} / Z_{\mathrm{g}}\right) / P_{\text{rated }} .
$$

A series of tests, shown in Table II, were conducted to isolate and confirm the roles of both the SST's internal characteristics and the external grid impedance as the key contributors to system instability.

\begin{table}[H]
\begin{center}
\caption{Definition of Experimental Cases for Stability Validation}
\begin{tabular}{|l|l|l|l|}
\hline
Case & Controller Configuration & DAB Status & Grid Inductance (SCR) \\
\hline
Case 1 & SOGI-PLL ( $k_{i \text{ sogi }}=1$ ) & Active & $3.8 \mathrm{mH}(2.62)$ \\
\hline
Case 2 & SOGI-PLL ( $k_{i \text{ sogi }}=1$ ) & Deactivated & $3.8 \mathrm{mH}(2.62)$ \\
\hline
Case 3 & SOGI-PLL ( $k_{i \text{ sogi }}=1$ ) & Active & $3.0 \mathrm{mH}(3.32)$ \\
\hline
Case 4 & SOGI-PLL ( $k_{i \text{ sogi }}=0.5$ ) & Active & $3.0 \mathrm{mH}(3.32)$ \\
\hline
Case 5 & Designed DF-OSR & Active & $3.8 \mathrm{mH}(2.62)$ \\
\hline
Case 6 & Designed DF-OSR & Active & $3.0 \mathrm{mH}(3.32)$ \\
\hline
\end{tabular}
\end{center}
\end{table}

\section*{B. Validation of the Predicted Instability Mechanisms}
This section presents experimental results designed to validate the instability mechanisms diagnosed by the dissipativity analysis in Section IV-A.

First, to verify that the active control of the DAB stage is the primary internal source of nonpassive behavior, an experiment was conducted as shown in Fig. 9(a). Here, the system response is compared under two conditions while connected to a weak grid: with the destabilizing dynamics of the DAB active (Case 1) and with them deactivated (Case 2, where the DAB is replaced by an equivalent passive impedance). As shown in Fig. 9(a), the system is initially stable in Case 2. Upon transitioning to Case 1, low-frequency oscillations immediately develop and grow in the input current (i) and the DC-link voltages $\left(v_{\mathrm{dc}}, v_{\mathrm{o}}\right)$. The zoomed-in view reveals an oscillation frequency of approximately 13 Hz . When the system is switched back to Case 2, the oscillations are rapidly damped, and stability is restored. This result confirms that the control strategy of the DAB converter introduces the intrinsic instability risk, which directly corresponds to the negative self-dissipativity property $P_{\mathrm{dc}}$ identified in our theoretical assessment.

To validate the diagnostic capability for AC-side instabilities, Case 3 and Case 4 were conducted by dynamically altering the SOGI gain of the controller. The results are presented in Fig. 9(b). The experiment begins in Case 3, where the SST operates stably under a strong grid with a nominal SOGI gain ( $k_{\text{isogi }}=1$ ). As shown, the input current ( $i_{\text{in }}$ ) and the DC voltages ( $v_{\mathrm{dc}}, v_{\mathrm{o}}$ ) are well-regulated with minimal distortion. Subsequently, the controller is transitioned to Case 4, where the SOGI gain is switched to a low value ( $k_{\text{isogi }}=0.5$ ), which was predicted by the dissipativity analysis to cause instability. As depicted in the right half of the figure, the system becomes unstable, exhibiting growing low-frequency oscillations. This result provides validation for the theoretical diagnosis, confirming that an improperly tuned SOGI gain leads to a failure in the AC-side self-dissipativity ( $P_{d q}<0$ ), thereby triggering instability. The stable transient performance of the baseline controller under Case 3 to DC voltage and load steps is presented in Fig. 9(c) and (d), respectively.

It is noteworthy to compare the experimental oscillation frequency with the theoretical prediction. The theoretical dissipativity analysis in Fig. 5 predicts that the system is most

\begin{figure}[H]
\begin{center}
  \includegraphics[alt={},max width=\columnwidth]{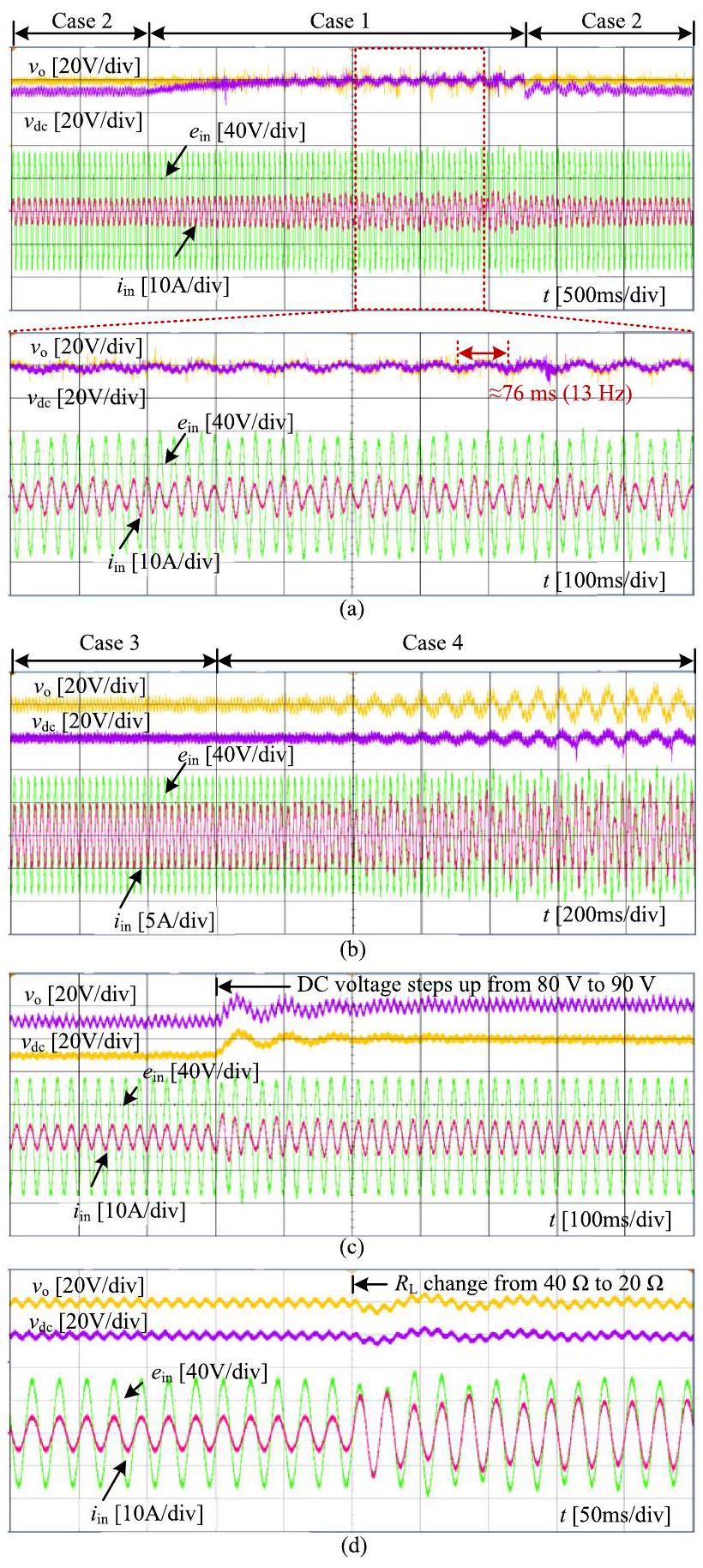}
\caption{Experimental validation of instability mechanisms. (a) Instability induced by the DC-side: low-frequency oscillations at 13 Hz emerge when the DAB active control is enabled (case 1). (b) Instability induced by the AC-side: oscillations emerge when the SOGI gain is switched from a nominal to a low value (case 4). (c) Transient response to a DC voltage step (case 3). (d) Transient response to a load step (case 3) (THD: 6.27\% $\rightarrow 7.02 \%$ ).}
\end{center}
\end{figure}

vulnerable to instability at 11 Hz , where the coupling indices exhibit the deepest negative dip. This prediction aligns well with the experimental results in Case 1, where the actual oscillation frequency is measured at approximately 13 Hz . This quantitative agreement confirms that the instability is indeed driven

\begin{figure}[H]
\begin{center}
  \includegraphics[alt={},max width=\columnwidth]{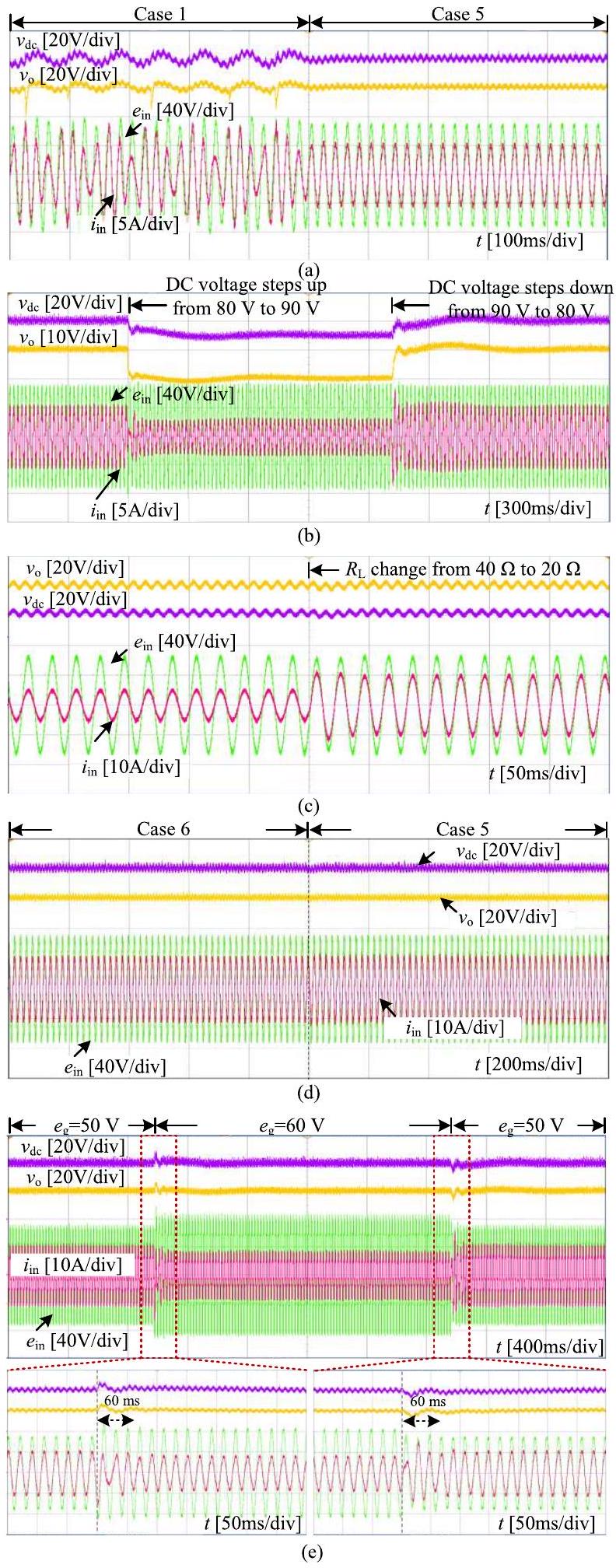}
\caption{Experimental validation of the enhanced controller. (a) System response demonstrating instability suppression when switching from the SOGI-PLL controller (case 1) to the proposed DF-OSR strategy (case 5). (b) Dynamic performance of the enhanced controller during a step change in DC operating voltage, demonstrating its robustness. (c) Its dynamic performance during a load step (THD: $6.23 \% \rightarrow 5.58 \%$ ). (d) System response with the DF-OSR controller operating under a various grid impedance. (e) System response with the DF-OSR controller operating under a various grid voltage.}
\end{center}
\end{figure}

\begin{table}[H]
\begin{center}
\caption{Comparative Analysis of Various Stability Enhancement Techniques}
\begin{tabular}{|l|l|l|l|}
\hline
Technique Category & Implementation Complexity & Parameter Tuning Effort & Dependence on System Model \\
\hline
Filter-based reshaping [33] & Low & High (requires critical frequency) & High (design criteria needed) \\
\hline
Adaptive/ AI-based [34] & High & Low (selflearning) & Low \\
\hline
Virtual impedance [35] & Medium & Medium (depends on grid estimation) & Medium \\
\hline
Proposed & Low & No additional parameters & Low \\
\hline
\end{tabular}
\end{center}
\end{table}

by the specific coupling dynamics identified by the proposed admittance. Additionally, it has been shown that system instability can be triggered by either the negative dissipativity of DCport ( $P_{\mathrm{dc}}<0$ ) or the internal dynamic deficiencies of AC-port ( $P_{d q}<0$ ).

\section*{C. Validation of the Enhancement Strategy and Its Robustness}
The effectiveness of the DF-OSR controller is demonstrated in Fig. 10(a). The experiment was initiated with the SOGIPLL controller operating under the same weak grid conditions ( $L_{\mathrm{g}}=3.8 \mathrm{mH}$ ) that previously led to instability (Case 1). As shown in the left half of the figure, the system exhibits the sustained low-frequency oscillation. Subsequently, the control algorithm was dynamically switched to the proposed DF-OSR enhancement strategy (Case 5). Upon activation, the oscillations in the input current ( $i_{\text{in }}$ ) and the DC-link voltages ( $v_{\text{dc }}, v_{\mathrm{o}}$ ) are immediately and effectively suppressed. The system rapidly settles into a stable, steady-state operation. This result provides definitive experimental proof that the proposed enhancement strategy successfully resolves the coupling-induced instability, directly validating the theoretical dissipativity re-assessment presented in Section IV-C.

To further evaluate the practical viability and robustness of the enhanced design, its performance during a operational transient was tested. Fig. 10(b) and (c) illustrate the dynamic response of the system to a step change in the DC operating voltage reference and a step change in load, respectively, while the DF-OSR controller is active. The controller demonstrates well dynamic characteristics, achieving a fast and well-damped transition to the new steady-state point without signs of oscillatory behavior.

Finally, the DF-OSR controller was tested under the stiffer grid condition (Case 6). Fig. 10(d) illustrates the system dynamics when the grid inductance steps from $L_{\mathrm{g}}=3.0 \mathrm{mH}$ (SCR $\approx$ 3.32) to $L_{\mathrm{g}}=3.8 \mathrm{mH}$ (SCR $\approx 2.62$ ). As observed, the system operates stably with low ripple, confirming no performance degradation under stiff conditions. Furthermore, the robustness of the proposed strategy against grid voltage fluctuations is validated as shown in Fig. 10(e). When the grid voltage $e_{\mathrm{g}}$\\
undergoes step changes between 50 and 60 V , the system maintains stable operation with fast voltage regulation and negligible current distortion, further verifying the dynamic performance of the enhanced controller.

To contextualize the advantages of the proposed DF-OSR strategy, Table III provides a comparison with several the state-of-the-art stability enhancement techniques used in SST. The proposed method offers a compelling alternative by eliminating the need for additional parameter tuning for the stabilization loop and reducing model dependency, thereby simplifying the design process while enhancing robustness.

\section*{VI. Conclusion}
This article presents a multiport dissipativity framework to diagnose and mitigate coupling-induced instabilities in singlephase SSTs. By leveraging the decomposition of the dissipativity criterion, the framework effectively distinguishes between localized and coupling-induced stability threats, revealing that the dominant instability mechanism is a coupling-dissipativity failure induced by the internal dynamics of the synchronization loop. Guided by this diagnosis, a DF-OSR enhancement strategy is proposed to resolve the coupling failure by eliminating the adverse dynamics of SOGI-based filters. This method simplifies the design process by introducing no additional tunable parameters, ensuring robust operation without complex optimization. The diagnostic accuracy and the effectiveness of the proposed strategy are validated through hardware experiments.

However, it is worth noting that while the proposed method is effective for stability enhancement under weak grid conditions and small-signal disturbances, it relies on the reference tracking assumption. Consequently, it may face challenges during severe grid-side faults where the actual current deviates significantly from the reference. In such large-signal fault scenarios, switching to a dedicated fault-ride-through control strategy would be required.

\section*{References}
[1] X. She, A. Q. Huang, and R. Burgos, "Review of solid-state transformer technologies and their application in power distribution systems," IEEE Trans. Emerg. Sel. Topics Power Electron., vol. 1, no. 3, pp. 186-198, Sep. 2013.\\[0pt]
[2] Q. Ye, R. Mo, and H. Li, "Impedance modeling and DC bus voltage stability assessment of a solid-state-transformer-enabled hybrid AC-DC grid considering bidirectional power flow," IEEE Trans. Ind. Electron., vol. 67, no. 8, pp. 6531-6540, Aug. 2020.\\[0pt]
[3] Y. Wang, X. Wang, F. Blaabjerg, and Z. Chen, "Harmonic instability assessment using state-space modeling and participation analysis in inverter-fed power systems," IEEE Trans. Ind. Electron., vol. 64, no. 1, pp. 806-816, Jan. 2017.\\[0pt]
[4] Y. Zhu, Y. Gu, Y. Li, and T. C. Green, "Impedance-based rootcause analysis: Comparative study of impedance models and calculation of eigenvalue sensitivity," IEEE Trans. Power Syst., vol. 38, no. 2, pp. 1642-1654, Mar. 2023.\\[0pt]
[5] Y. Liao, Z. Liu, H. Zhang, and B. Wen, "Low-frequency stability analysis of single-phase system with $d q$-frame impedance approachpart i: Impedance modeling and verification," IEEE Trans. Ind. Appl., vol. 54, no. 5, pp. 4999-5011, Sep./Oct. 2018.\\[0pt]
[6] J. Sun, "Impedance-based stability criterion for grid-connected inverters," IEEE Trans. Power Electron., vol. 26, no. 11, pp. 3075-3078, Nov. 2011.\\[0pt]
[7] J. Sun, "Frequency-domain stability criteria for converter-based power systems," IEEE Open J. Power Electron., vol. 3, pp. 222-254, 2022.\\[0pt]
[8] J. Wyatt, L. Chua, J. Gannett, I. Goknar, and D. Green, "Energy concepts in the state-space theory of nonlinear n-ports: Part I-passivity," IEEE Trans. Circuits Syst., vol. 28, no. 1, pp. 48-61, Jan. 1981.\\[0pt]
[9] L. Harnefors, X. Wang, A. G. Yepes, and F. Blaabjerg, "Passivitybased stability assessment of grid-connected VSCs-an overview," IEEE Trans. Emerg. Sel. Topics Power Electron., vol. 4, no. 1, pp. 116125, Mar. 2016.\\[0pt]
[10] Fixed installations and rolling stock for railway applications - Technical criteria for the coordination between electric traction power supply systems and rolling stock to achieve interoperability - Part 2: Stability and harmonics, European Committee for Electrotechnical Standardization (CENELEC) European Standard EN 50 388-2:2025, Mar. 2025.\\[0pt]
[11] L. Harnefors, X. Wang, S.-F. Chou, M. Bongiorno, M. Hinkkanen, and M. Routimo, "Asymmetric complex-vector models with application to VSC-grid interaction," IEEE Trans. Emerg. Sel. Topics Power Electron., vol. 8, no. 2, pp. 1911-1921, Jun. 2020.\\[0pt]
[12] M. Li, E. Liu, H. Geng, Y. Mao, X. Wang, and X. Zhang, "Passivitybased control for the stability of grid-forming multi-inverter power stations," IEEE Trans. Ind. Electron., vol. 72, no. 9, pp. 9117-9127, Sep. 2025.\\[0pt]
[13] F. Chen et al., "Limitations of using passivity index to analyze gridinverter interactions," IEEE Trans. Power Electron., vol. 39, no. 11, pp. 14465-14477, Nov. 2024.\\[0pt]
[14] F. Chen et al., "An extended frequency-domain passivity theory for MIMO dynamics specifications of voltage-source inverters," IEEE Trans. Power Electron, vol. 40, no. 2, pp. 2943-2957, Feb. 2025.\\[0pt]
[15] R. Cvetanović, I. Z. Petrić, P. Mattavelli, and S. Buso, "All-port MIMO admittance passivity for robust stability of DC-DC interlinking converters," IEEE Trans. Power Electron., vol. 39, no. 10, pp. 1299113008, Oct. 2024.\\[0pt]
[16] R. Cvetanović, I. Z. Petrić, P. Mattavelli, and S. Buso, "All-port unterminated admittance passivity for robust stability of AC-DC interlinking converters," IEEE Trans. Power Electron., vol. 40, no. 9, pp. 1388013894, Sep. 2025.\\[0pt]
[17] J. Sun, "Two-port characterization and transfer immittances of AC-DC converters-part i: Modeling," IEEE Open J. Power Electron., vol. 2, pp. 440-462, 2021.\\[0pt]
[18] J. Pedra, L. Sainz, and L. Monjo, "Three-port small signal admittancebased model of VSCs for studies of multi-terminal HVDC hybrid AC/DC transmission grids," IEEE Trans. Power Syst., vol. 36, no. 1, pp. 732-743, Jan. 2021.\\[0pt]
[19] I. Marzo, A. Sanchez-Ruiz, J. A. Barrena, G. Abad, and I. Muguruza, "Power balancing in cascaded H -bridge and modular multilevel converters under unbalanced operation: A review," IEEE Access, vol. 9, pp. 110525-110543, 2021.\\[0pt]
[20] C. Gu, Z. Zheng, L. Xu, K. Wang, and Y. Li, "Modeling and control of a multiport power electronic transformer (PET) for electric traction applications," IEEE Trans. Power Electron., vol. 31, no. 2, pp. 915-927, Feb. 2016.\\[0pt]
[21] Y. Hong, Y. Li, Z. Shuai, and H. Yang, "Low-frequency stability analysis of power electronic traction transformer based train-network system," CSEE J. Power Energy Syst., vol. 10, no. 4, pp. 1608-1617, Jul. 2024.\\[0pt]
[22] X. She, A. Q. Huang, T. Zhao, and G. Wang, "Coupling effect reduction of a voltage-balancing controller in single-phase cascaded multilevel converters," IEEE Trans. Power Electron., vol. 27, no. 8, pp. 35303543, Aug. 2012.\\[0pt]
[23] J. Bao and P. L. Lee, Process Control: The Passive Systems Approach. London, U.K.: Springer-Verlag, 2007.\\[0pt]
[24] G. T. Gilbert, "Positive definite matrices and Sylvester's criterion," Amer. Math. Monthly, vol. 98, no. 1, pp. 44-46, 1991.\\[0pt]
[25] M. A. Hassan, C.-L. Su, F.-Z. Chen, and K.-Y. Lo, "Adaptive passivitybased control of a DC-DC boost power converter supplying constant power and constant voltage loads," IEEE Trans. Ind. Electron., vol. 69, no. 6, pp. 6204-6214, Jun. 2022.\\[0pt]
[26] I. Z. Petric, P. Mattavelli, and S. Buso, "Passivation of grid-following VSCs: A comparison between active damping and multi-sampled PWM," IEEE Trans. Power Electron., vol. 37, no. 11, pp. 13205-13216, Nov. 2022.\\[0pt]
[27] S. Li, W. Qi, S.-C. Tan, and S. Hui, "Enhanced automatic-powerdecoupling control method for single-phase AC-to-DC converters," IEEE Trans. Power Electron., vol. 33, no. 2, pp. 1816-1828, Feb. 2018.\\[0pt]
[28] J. He, L. Du, B. Liang, Y. Li, and C. Wang, "A coupled virtual impedance for parallel AC/DC converter based power electronics system," IEEE Trans. Smart Grid, vol. 10, no. 3, pp. 3387-3400, May 2019.\\[0pt]
[29] F. Blaabjerg, Control of Power Electronic Converters and Systems: volume 2, Academic Press, 2018. vol. 2.\\[0pt]
[30] S. Golestan, M. Monfared, F. D. Freijedo, and J. M. Guerrero, "Dynamics assessment of advanced single-phase PLL structures," IEEE Trans. Ind. Electron., vol. 60, no. 6, pp. 2167-2177, Jun. 2013.\\[0pt]
[31] H. Zong, C. Zhang, X. Cai, and M. Molinas, "Three-port impedance model and validation of vscs for stability analysis," in Proc. Annu. Meeting CSEE Study Comm. HVDC Power Electron. (HVDC), vol. 2021. IET, 2021, pp. 293-298.\\[0pt]
[32] "IEEE standard for harmonic control in electric power systems," New York, NY, USA, Aug. 2022, IEEE Std 519-2022.\\[0pt]
[33] Z. Zou, J. Tang, S. Yuan, G. Buticchi, and M. Liserre, "A generalized design criterion of filter-based stabilizing control of ST-Fed LV grid," IEEE Trans. Emerg. Sel. Topics Power Electron., vol. 13, no. 2, pp. 2258-2269, Apr. 2025.\\[0pt]
[34] J. Tang, Z. Zou, J. Yang, G. Buticchi, and W. Hua, "Adaptive stabilizing control of smart transformer based on reinforcement learning optimization," IEEE Trans. Ind. Appl., vol. 60, no. 3, pp. 4324-4337, May/Jun. 2024.\\[0pt]
[35] V. R. Chowdhury, R. P. Kandula, and D. Divan, "Negative virtual inductance based active damping and direct power control of a soft switching solid state transformer for $p v$ application," in Proc. IEEE Appl. Power Electron. Conf. Expo. (APEC), Piscataway, NJ, USA: IEEE Press, 2022, pp. 1950-1955.\\

\end{document}